\documentclass[12pt,prd,showpacs,tightenlines,nofootinbib]{revtex4}
\usepackage{bm}
\usepackage{graphics}
\usepackage{rotating}
\usepackage{epsfig}
\begin{document}
\begin{flushright}{HU-EP-07/21}\\HEPHY-PUB 842/07\end{flushright}
\title{Masses of tetraquarks with two heavy quarks in the
relativistic quark model}
\author{D.~Ebert}
\affiliation{Institut f\"ur Physik, Humboldt--Universit\"at zu
Berlin, Newtonstr.~15, D-12489 Berlin, Germany}
\author{R.~N.~Faustov}
\author{V.~O.~Galkin}
\affiliation{Institut f\"ur Physik, Humboldt--Universit\"at zu
Berlin, Newtonstr.~15, D-12489 Berlin, Germany}
\affiliation{Dorodnicyn Computing Centre, Russian Academy of
Sciences, Vavilov Str.~40, 119991 Moscow, Russia}
\author{W.~Lucha}
\affiliation{Institute for High Energy Physics, Austrian Academy of
 Sciences, Nikolsdorfergasse 18, A-1050, Vienna, Austria}
\begin{abstract}
Masses of tetraquarks with two heavy quarks and open charm and
bottom are calculated in the framework of the diquark-antidiquark
picture in the relativistic~quark model. All model parameters were
regarded as fixed by previous considerations~of various properties
of mesons and baryons. The light quarks and diquarks are treated
completely relativistically. 
The $c$ quark is assumed to be heavy enough to make the diquark
configurations dominating. 
The diquarks are considered not to be 
point-like but to have an internal structure which is taken into
account by the calculated diquark form factor entering the
diquark-gluon interaction. It is found that all the $(cc)(\bar
q\bar q')$ tetraquarks have masses above the thresholds for decays
into open charm mesons. Only the $I(J^P)=0(1^+)$ state of
$(bb)(\bar u\bar d)$ lies below the $BB^*$ threshold and is
predicted to be narrow.
\end{abstract}


\maketitle

\section{Introduction}
\label{sec:intr}

Recent experimental studies of the heavy meson spectroscopy
revealed several new states, such as $X(3872)$, $Y(4260)$,
$D^*_{s0}(2317)$ etc., which cannot be simply accommodated in the
quark-antiquark ($q\bar q$) picture \cite{swanson}. These states
can be considered as indications of the possible existence of
exotic multiquark states which were proposed long ago,
e.g.~in~\cite{jaffe}. The idea to revisit the multiquark picture
using diquarks has been put forward by Jaffe and Wilczek
\cite{jw}. At present, vast experimental and theoretical evidence
of the important role played by diquark correlations in hadrons is
collected \cite{jpr}.

The simplest multiquark system is a tetraquark, composed of two
quarks and two antiquarks. Heavy tetraquarks are of particular
interest, since the presence of a heavy quark increases the
binding energy of the bound system and, as a result, the
possibility that such tetraquarks will have masses below the
thresholds for decays to mesons with open heavy flavour. If such
strong decays are kinematically forbidden, then corresponding
tetraquarks can decay only weakly or electromagnetically and thus
they should have a small decay width. In this paper we consider
tetraquarks with two heavy quarks as bound systems of a diquark
and antidiquark.
Therefore we assume that both $b$ and $c$ quarks are heavy enough to
make the attractive $QQ$ interaction stronger than the $Q\bar q$ one. For
tetraquarks containing $c$ quarks the obtained results crucially
depend on this assumption.  
In particular, the doubly heavy $(QQ')(\bar q\bar
q')$ tetraquark ($Q=b,c$ and $q=u,d,s$) is considered as the bound
system of the heavy diquark ($QQ'$) and light antidiquark ($\bar
q\bar q'$), while the $(cq)(\bar b\bar q')$ tetraquark is the
bound state of the heavy-light diquark ($cq$) and antidiquark
($\bar b\bar q')$. Masses of heavy tetraquarks with hidden charm
($cq)(\bar c \bar q)$ and bottom $(bq)(\bar b\bar q)$ were
calculated in our previous paper \cite{efght}. There the dynamical
analysis has shown that $X(3872)$ and $Y(4260)$ can be indeed the
diquark-antidiquark tetraquarks with hidden charm. It was also
argued that the corresponding ground-state tetraquarks with hidden
bottom have masses below the open bottom threshold, and thus they
should be narrow states.

It is important to investigate the possible stability of the
$(QQ')(\bar q\bar q')$ tetraquarks since they are explicitly
exotic states with the heavy flavour number equal to 2. Thus,
their observation would be a direct proof of the existence of the
multiquark states. Estimates of the production rates of such
tetraquarks indicate that they could be produced and detected at
present (SELEX, Tevatron, RHIC) and future facilities (LHC, LHCb,
ALICE) \cite{fjrt}.

To calculate the masses of heavy tetraquarks we use the
relativistic quark model based on the quasipotential approach in
quantum field theory. Previously we considered in our model the
mass spectra of the ground-state and excited doubly-heavy ($QQq$)
\cite{efgm} and heavy ($qqQ$) \cite{hbar,exhbar} baryons in the
heavy-diquark--light-quark and light-diquark--heavy-quark
approximations, respectively. The light quarks and light diquarks
were treated completely relativistically. The internal structure
of the light and heavy diquarks was taken into account by
calculating diquark-gluon form factors in terms of the obtained
diquark wave functions. Such scheme proved to be very effective
and successful in our calculation of the masses of heavy baryons
in good agreement with experimental data \cite{pdg}. The predicted
masses of the $\Omega_c^*$, $\Sigma_b$, $\Sigma_b^*$ and $\Xi_b$
baryons proved to be very close to the recently measured ones
\cite{babaromega,cdfSigma,Xib}. Moreover, in Ref.~\cite{exhbar} it
was shown that currently available experimental data on excited
charmed baryons can be accommodated in the picture treating a
heavy baryon as the bound system of the light diquark and heavy
quark, experiencing orbital and radial excitations. This gives us
additional confidence in the reliability of the diquark
approximation within our model and motivates the consideration of
tetraquarks as diquark-antidiquark bound systems. It is important
to note that all parameters of our model were fixed in the
previous calculations of meson mass spectra and decays, and we
will keep their values in the following analysis of heavy
tetraquarks.

\section{Relativistic quark model}
\label{sec:rqm}

In the quasipotential approach and diquark-antidiquark picture of
heavy tetraquarks the interaction of two quarks in a diquark and
the diquark-antidiquark interaction in a tetraquark are described
by the diquark wave function ($\Psi_{d}$) of the bound quark-quark
state and by the tetraquark wave function ($\Psi_{T}$) of the
bound diquark-antidiquark state, respectively, which satisfy the
quasipotential equation of the Schr\"odinger type \cite{efg}
\begin{equation}
\label{quas}
{\left(\frac{b^2(M)}{2\mu_{R}}-\frac{{\bf
p}^2}{2\mu_{R}}\right)\Psi_{d,T}({\bf p})} =\int\frac{d^3 q}{(2\pi)^3}
 V({\bf p,q};M)\Psi_{d,T}({\bf q}),
\end{equation}
where the relativistic reduced mass is
\begin{equation}
\mu_{R}=\frac{E_1E_2}{E_1+E_2}=\frac{M^4-(m^2_1-m^2_2)^2}{4M^3},
\end{equation}
and $E_1$, $E_2$ are given by
\begin{equation}
\label{ee}
E_1=\frac{M^2-m_2^2+m_1^2}{2M}, \quad E_2=\frac{M^2-m_1^2+m_2^2}{2M}.
\end{equation}
Here, $M=E_1+E_2$ is the bound-state mass (diquark or tetraquark),
$m_{1,2}$ are the masses of quarks ($q_1$ and $q_2$) which form
the diquark or of the diquark ($d$) and antiquark ($d'$) which
form the heavy tetraquark ($T$), and ${\bf p}$ is their relative
momentum. In the center-of-mass system the relative momentum
squared on mass shell reads
\begin{equation}
{b^2(M) }
=\frac{[M^2-(m_1+m_2)^2][M^2-(m_1-m_2)^2]}{4M^2}.
\end{equation}

The kernel $V({\bf p,q};M)$ in Eq.~(\ref{quas}) is the
quasipotential operator of the quark-quark or diquark-antidiquark
interaction. It is constructed with the help of the off-mass-shell
scattering amplitude, projected onto the positive-energy states.
In the following analysis we closely follow the similar
construction of the quark-antiquark interaction in mesons which
were extensively studied in our relativistic quark model
\cite{efg,egf}. For the quark-quark interaction in a diquark we
use the relation $V_{qq}=V_{q\bar q}/2$ arising under the
assumption of an octet structure of the interaction from the
difference in the $qq$ and $q\bar q$ colour
states.\footnote{Obviously, it is important to study diquark
correlations in gauge-invariant color-singlet hadron states on the
lattice.} An important role in this construction is played by the
Lorentz structure of the confining interaction. In our analysis of
mesons, while constructing the quasipotential of the
quark-antiquark interaction, we assumed that the effective
interaction is the sum of the usual one-gluon exchange term and a
mixture of long-range vector and scalar linear confining
potentials, where the vector confining potential contains the
Pauli terms. We use the same conventions for the construction of
the quark-quark and diquark-antidiquark interactions in the
tetraquark. The quasipotential is then defined as follows
\cite{efgm,egf}.

(a) For the quark-quark ($qq'$), ($Qq$), ($QQ'$) interactions,
$V({\bf p,q};M)$ reads
 \begin{equation}
\label{qpot}
V({\bf p,q};M)=\bar{u}_{1}(p)\bar{u}_{2}(-p){\cal V}({\bf p}, {\bf
q};M)u_{1}(q)u_{2}(-q),
\end{equation}
with
\[
{\cal V}({\bf p,q};M)=\frac12\left[\frac43\alpha_sD_{ \mu\nu}({\bf
k})\gamma_1^{\mu}\gamma_2^{\nu}+ V^V_{\rm conf}({\bf k})
\Gamma_1^{\mu}({\bf k})\Gamma_{2;\mu}(-{\bf k})+
 V^S_{\rm conf}({\bf k})\right].
\]
Here, $\alpha_s$ is the QCD coupling constant; $D_{\mu\nu}$ is the
gluon propagator in the Coulomb gauge,
\begin{equation}
D^{00}({\bf k})=-\frac{4\pi}{{\bf k}^2}, \quad D^{ij}({\bf k})=
-\frac{4\pi}{k^2}\left(\delta^{ij}-\frac{k^ik^j}{{\bf k}^2}\right),
\quad D^{0i}=D^{i0}=0,
\end{equation}
and ${\bf k=p-q}$; $\gamma_{\mu}$ and $u(p)$ are the Dirac
matrices and spinors,
\begin{equation}
\label{spinor}
u^\lambda({p})=\sqrt{\frac{\epsilon(p)+m}{2\epsilon(p)}}
\left(\begin{array}{c} 1\\
\displaystyle\frac{\mathstrut\bm{\sigma}\cdot{\bf p}}
{\mathstrut\epsilon(p)+m}
\end{array}\right)
\chi^\lambda,
\end{equation}
with $\epsilon(p)=\sqrt{{\bf p}^2+m^2}$.

The effective long-range vector vertex of the quark is
defined \cite{egf} by
\begin{equation}
\Gamma_{\mu}({\bf k})=\gamma_{\mu}+
\frac{i\kappa}{2m}\sigma_{\mu\nu}\tilde k^{\nu}, \qquad \tilde
k=(0,{\bf k}),
\end{equation}
where $\kappa$ is the Pauli interaction constant characterizing
the anomalous chromomagnetic moment of quarks. In configuration
space the vector and scalar confining potentials in the
nonrelativistic limit reduce to
\begin{eqnarray}
V^V_{\rm conf}(r)&=&(1-\varepsilon)V_{\rm
conf}(r),\nonumber\\[1ex] V^S_{\rm conf}(r)& =&\varepsilon V_{\rm
conf}(r),
\end{eqnarray}
with
\begin{equation}
V_{\rm conf}(r)=V^S_{\rm conf}(r)+
V^V_{\rm conf}(r)=Ar+B,
\end{equation}
where $\varepsilon$ is the mixing coefficient.

(b) For the diquark-antidiquark ($d\bar d'$) interaction, $V({\bf
p,q};M)$ is given by
\begin{eqnarray}
\label{dpot} V({\bf p,q};M)&=&\frac{\langle
d(P)|J_{\mu}|d(Q)\rangle} {2\sqrt{E_dE_d}} \frac43\alpha_sD^{
\mu\nu}({\bf k})\frac{\langle d'(P')|J_{\nu}|d'(Q')\rangle}
{2\sqrt{E_{d'}E_{d'}}}\nonumber\\[1ex]
&&+\psi^*_d(P)\psi^*_{d'}(P')\left[J_{d;\mu}J_{d'}^{\mu} V_{\rm
conf}^V({\bf k})+V^S_{\rm conf}({\bf
k})\right]\psi_d(Q)\psi_{d'}(Q'),
\end{eqnarray}
where $\langle
d(P)|J_{\mu}|d(Q)\rangle$ is the vertex of the
diquark-gluon interaction which takes into account the finite size of
the diquark and is discussed
below
$\Big[$$P^{(')}=(E_{d^{(')}},\pm{\bf p})$ and
$Q^{(')}=(E_{d^{(')}},\pm{\bf q})$,
$E_d=(M^2-M_{d'}^2+M_d^2)/(2M)$ and $E_{d'}=(M^2-M_d^2+M_{d'}^2)/(2M)$
$\Big]$.

The diquark state in the confining part of the diquark-antidiquark
quasipotential (\ref{dpot}) is described by the wave functions
\begin{equation}
 \label{eq:ps}
 \psi_d(p)=\left\{\begin{array}{ll}1 &\qquad \text{for a scalar
 diquark,}\\[1ex]
\varepsilon_d(p) &\qquad \text{for an axial-vector diquark,}
\end{array}\right.
\end{equation}
where the four-vector
\begin{equation}\label{pv}
\varepsilon_d(p)=\left(\frac{(\bm{\varepsilon}_d\cdot{\bf
p})}{M_d},\bm{\varepsilon}_d+ \frac{(\bm{\varepsilon}_d\cdot{\bf
p}){\bf
 p}}{M_d(E_d(p)+M_d)}\right), \qquad \varepsilon^\mu_d(p) p_\mu=0,
\end{equation}
is the polarization vector of the axial-vector diquark with
momentum ${\bf p}$, $E_d(p)=\sqrt{{\bf p}^2+M_d^2}$, and
$\varepsilon_d(0)=(0,\bm{\varepsilon}_d)$ is the polarization
vector in the diquark rest frame. The effective long-range vector
vertex of the diquark can be presented in the form
\begin{equation}
 \label{eq:jc}
 J_{d;\mu}=\left\{\begin{array}{ll}
 \frac{\displaystyle (P+Q)_\mu}{\displaystyle
 2\sqrt{E_dE_d}}&\qquad \text{ for a scalar diquark,}\\[3ex]
-\; \frac{\displaystyle (P+Q)_\mu}{\displaystyle2\sqrt{E_dE_d}}
 +\frac{\displaystyle i\mu_d}{\displaystyle 2M_d}\Sigma_\mu^\nu
\tilde k_\nu
 &\qquad \text{ for an axial-vector diquark,}\end{array}\right.
\end{equation}
where $\tilde k=(0,{\bf k})$. Here, the antisymmetric tensor
$\Sigma_\mu^\nu$ is defined by
\begin{equation}
 \label{eq:Sig}
 \left(\Sigma_{\rho\sigma}\right)_\mu^\nu=-i(g_{\mu\rho}\delta^\nu_\sigma
 -g_{\mu\sigma}\delta^\nu_\rho),
\end{equation}
and the axial-vector diquark spin ${\bf S}_d$ is given by
$(S_{d;k})_{il}=-i\varepsilon_{kil}$; $\mu_d$ is the total
chromomagnetic moment of the axial-vector diquark.

The constituent quark masses $m_b=4.88$ GeV, $m_c=1.55$ GeV,
$m_u=m_d=0.33$ GeV, $m_s=0.5$ GeV and the parameters of the linear
potential $A=0.18$ GeV$^2$ and $B=-0.3$~GeV have the values
typical in quark models. The value of the mixing coefficient of
vector and scalar confining potentials $\varepsilon=-1$ has been
determined from the consideration of charmonium radiative decays
\cite{efg} and the heavy-quark expansion \cite{fg}. The universal
Pauli interaction constant $\kappa=-1$ has been fixed from the
analysis of the fine splitting of heavy quarkonia ${ }^3P_J$-
states \cite{efg}. In this case, the long-range chromomagnetic
interaction of quarks vanishes in accordance with the flux-tube
model.

\section{Light, heavy-light and heavy diquarks}
\label{sec:lhd}

As the first step, we calculate the masses and form factors of the
diquarks. As is well known, the light quarks are highly
relativistic, which makes the $v/c$ expansion inapplicable and
thus a completely relativistic treatment is required. To achieve
this goal in describing light and heavy-light diquarks, we closely
follow our recent consideration of the spectra of light mesons and
adopt the same procedure to make the relativistic quark potential
local by replacing $\epsilon_{1,2}(p)\equiv\sqrt{m_{1,2}^2+{\bf
p}^2}$ by $E_{1,2}$ (see the discussion in Ref.~\cite{lmes}). The
resulting quark--quark interaction potential is equal to 1/2 of
the $q\bar q$ interaction in the meson. We solve numerically the
quasipotential equation with this complete relativistic potential
which depends on the diquark mass in a complicated, highly
nonlinear way. The obtained ground-state masses of scalar and
axial-vector light and heavy diquarks \cite{hbar,efght,efgm} are
presented in Tables~\ref{tab:ldmass} and \ref{tab:dmass}.

\begin{table}
 \caption{Masses $M$ and form factor parameters (for definitions see
 Eq.~(\ref{eq:fr})) of light diquarks. $S$ and $A$
 denote scalar and axial-vector diquarks, antisymmetric $[q,q']$ and
 symmetric $\{q,q'\}$ in flavour, respectively. }
 \label{tab:ldmass}
\begin{ruledtabular}
\begin{tabular}{ccccc}
Quark content& Diquark type &
 $M$ (MeV)&$\xi$ (GeV)&$\zeta$ (GeV$^2$) \\
\hline $[u,d]$& S & 710 & 1.09 & 0.185 \\ $\{u,d\}$& A & 909 &
1.185 & 0.365 \\ $[u,s]$ & S& 948 & 1.23 & 0.225 \\ $\{u,s\}$& A &
1069 & 1.15 & 0.325 \\ $\{s,s\}$& A & 1203 & 1.13 & 0.280
\end{tabular}
\end{ruledtabular}
\end{table}

\begin{table}
 \caption{Masses $M$ and form factor parameters (for the definitions see
 Eq.~(\ref{eq:fr})) of heavy-light and heavy diquarks. $S$ and $A$
 denote scalar and axial-vector diquarks, antisymmetric $[Q,q]$ and
 symmetric $\{Q,q\}$ in flavour, respectively. }
 \label{tab:dmass}
\begin{ruledtabular}
\begin{tabular}{cccccccc}
Quark& Diquark&
\multicolumn{3}{l}{\underline{\hspace{2.5cm}$Q=c$\hspace{2.5cm}}}
\hspace{-3.4cm}
&\multicolumn{3}{l}{\underline{\hspace{2.5cm}$Q=b$\hspace{2.5cm}}}
\hspace{-3.4cm} \\ content &type & $M$ (MeV)&$\xi$ (GeV)&$\zeta$
(GeV$^2$) & $M$ (MeV)&$\xi$ (GeV)&$\zeta$ (GeV$^2$) \\ \hline
$[Q,u]$& $S$ & 1973& 2.55 &0.63 & 5359 &6.10 & 0.55 \\ $\{Q,u\}$&
$A$ & 2036& 2.51 &0.45 & 5381& 6.05 &0.35 \\ $[Q,s]$ & $S$& 2091&
2.15 & 1.05 & 5462 & 5.70 &0.35 \\ $\{Q,s\}$& $A$ & 2158&2.12&
0.99 & 5482 & 5.65 &0.27\\ $[Q,c]$ & $S$& & & & 6519 & 1.50
&0.59\\ $\{Q,c\}$& $A$& 3226& 1.30& 0.42 & 6526 & 1.50 &0.59\\
$\{Q,b\}$& $A$& 6526 & 1.50 &0.59& 9778 & 1.30 &1.60
 \end{tabular}
\end{ruledtabular}
\end{table}

In order to determine the diquark interaction with the gluon
field, which takes into account the diquark structure, it is
necessary to calculate the corresponding matrix element of the
quark current between diquark states. This (diagonal) matrix
element can be parameterized by the following set of elastic form
factors:

(a) scalar diquark ($S$)
\begin{equation}
 \label{eq:sff}
 \langle S(P)\vert J_\mu \vert S(Q)\rangle=h_+(k^2)(P+Q)_\mu,
\end{equation}

(b) axial-vector diquark ($A$)
\begin{eqnarray}
 \label{eq:avff}
\langle A(P)\vert J_\mu \vert A(Q)\rangle&=&
-[\varepsilon_d^*(P)\cdot\varepsilon_d(Q)]h_1(k^2)(P+Q)_\mu\nonumber\\[1ex]
&&+h_2(k^2) \left\{[\varepsilon_d^*(P) \cdot
Q]\varepsilon_{d;\mu}(Q)+
 [\varepsilon_d(Q) \cdot P]
\varepsilon^*_{d;\mu}(P)\right\}\nonumber\\[1ex]
&&+h_3(k^2)\frac1{M_{A}^2}[\varepsilon^*_d(P) \cdot Q]
 [\varepsilon_d(Q) \cdot P](P+Q)_\mu,
\end{eqnarray}
where $k=P-Q$, and $\varepsilon_d(P)$ is the polarization vector
of the axial-vector diquark (\ref{pv}).

Using the quasipotential approach with the impulse approximation for the
vertex function of the quark-gluon interaction, we find \cite{hbar}
\begin{eqnarray*}
 h_+(k^2)&=&h_1(k^2)=h_2(k^2)=F({\bf k}^2),\nonumber\\[1ex]
h_3(k^2)&=&0,
\end{eqnarray*}
\begin{eqnarray}\label{eq:hf}
F({\bf k}^2)&=&\frac{\sqrt{E_{d}M_{d}}}{E_{d}+M_{d}}
 \int \frac{d^3p}{(2\pi )^3} \bar\Psi_{d}
\left({\bf p}+ \frac{2\epsilon_{2}(p)}{E_{d}+M_{d}}{\bf k }
\right) \sqrt{\frac{\epsilon_1(p)+m_1}{\epsilon_1(p+k)+m_1}}
\Biggl[\frac{\epsilon_1(p+k)+\epsilon_1(p)}
{2\sqrt{\epsilon_1(p+k)\epsilon_1(p)}}\nonumber\\[1ex]
&&+\frac{{\bf p}\cdot{\bf k}}{2\sqrt{\epsilon_1(p+k)\epsilon_1(p)}
(\epsilon_1(p)+m_1)} \Biggr]\Psi_{d}({\bf
 p})+(1\longleftrightarrow 2),
\end{eqnarray}
where $\Psi_{d}$ are the diquark wave functions. We calculated the
corresponding form factors $F(r)/r$, which are the Fourier
transforms of $F({\bf k}^2)/{\bf k}^2$, using the diquark wave
functions found by numerically solving the quasipotential
equation. Our estimates show that this form factor can be
approximated with high accuracy by the expression
\begin{equation}
 \label{eq:fr}
 F(r)=1-e^{-\xi r -\zeta r^2},
\end{equation}
which agrees with previously used approximations \cite{efgm}. The
values of the parameters $\xi$ and $\zeta$ for light, heavy-light
and heavy scalar diquark $[q,q']$ and axial-vector diquark
$\{q,q'\}$ ground states are given in Tables~\ref{tab:ldmass} and
\ref{tab:dmass}.

\section{Masses of heavy tetraquarks}
\label{sec:mht}

As the second step, we calculate the masses of heavy tetraquarks
considered as bound states of diquark and antidiquark. For the
potential of the diquark-antidiquark interaction (\ref{dpot}) we
get \footnote{In our paper \cite{efght} first two spin-orbit terms
  were missed. However they do not influence published numerical results, since
  masses mostly of ground states were calculated. Orbital excitations
  were considered only for the tetraquarks composed of the scalar
  diquark and scalar antidiquark for which the missed terms vanish.}   
\begin{eqnarray}
 \label{eq:pot}
 V(r)&=& \hat V_{\rm Coul}(r)+V_{\rm conf}(r)+\frac12\Biggl\{\left[
   \frac1{E_1(E_1+M_1)}+\frac1{E_2(E_2+M_2)}\right]
\frac{\hat V'_{\rm Coul}(r)}r 
-\Biggl[\frac1{M_1(E_1+M_1)}\nonumber\\[1ex]
&& +\frac1{M_2(E_2+M_2)}\Biggr]
\frac{V'_{\rm conf}(r)}r +\frac{\mu_d}2
\left(\frac1{M_1^2}+\frac1{M_2^2}\right)\frac{V'^V_{\rm conf}(r)}r\Biggr\}
{\bf L}\cdot ({\bf
S}_1+{\bf S}_2 )\nonumber\\[1ex]
&&+\frac12\Biggl\{\left[
   \frac1{E_1(E_1+M_1)}-\frac1{E_2(E_2+M_2)}\right]
\frac{\hat V'_{\rm Coul}(r)}r 
-\left[\frac1{M_1(E_1+M_1)}-\frac1{M_2(E_2+M_2)}\right]\nonumber\\[1ex]
&& \times
\frac{V'_{\rm conf}(r)}r +\frac{\mu_d}2
\left(\frac1{M_1^2}-\frac1{M_2^2}\right)\frac{V'^V_{\rm conf}(r)}r\Biggl\}
{\bf L}\cdot ({\bf
S}_1-{\bf S}_2 )\nonumber\\[1ex]
&&+\frac1{E_1E_2}\Biggl\{{\bf
 p}\left[\hat V_{\rm Coul}(r)+V^V_{\rm conf}(r)\right]{\bf p} -\frac14
\Delta V^V_{\rm conf}(r)+ \hat V'_{\rm Coul}(r)\frac{{\bf
 L}^2}{2r}\nonumber\\[1ex]
&& +\frac1{r}\left[\hat V'_{\rm
Coul}(r)+\frac{\mu_d}4\left(\frac{E_1}{M_1}
+\frac{E_2}{M_2}\right)V'^V_{\rm conf}(r)\right]{\bf L}\cdot ({\bf
S}_1+{\bf S}_2)\nonumber\\[1ex] &&
+\frac{\mu_d}4\left(\frac{E_1}{M_1}
-\frac{E_2}{M_2}\right)\frac{V'^V_{\rm conf}(r)}{r}{\bf
L}\cdot({\bf S}_1-{\bf S}_2)\nonumber\\[1ex] &&
+\frac13\left[\frac1{r}{\hat V'_{\rm Coul}(r)}-\hat V''_{\rm
Coul}(r) +\frac{\mu_d^2}4\frac{E_1E_2}{M_1M_2}
\left(\frac1{r}{V'^V_{\rm conf}(r)}-V''^V_{\rm
 conf}(r)\right)\right]\nonumber\\[1ex]
&&\times
 \left[\frac3{r^2}({\bf S}_1\cdot{\bf r}) ({\bf
 S}_2\cdot{\bf r})-
{\bf S}_1\cdot{\bf S}_2\right]\nonumber\\[1ex] &&
+\frac23\left[\Delta \hat V_{\rm
Coul}(r)+\frac{\mu_d^2}4\frac{E_1E_2}{M_1M_2} \Delta V^V_{\rm
conf}(r)\right]{\bf S}_1\cdot{\bf S}_2\Biggr\},
\end{eqnarray}
where $$\hat V_{\rm Coul}(r)=-\frac{4}{3}\alpha_s
\frac{F_1(r)F_2(r)}{r}$$ is the Coulomb-like one-gluon exchange
potential which takes into account the finite sizes of the diquark
and antidiquark through corresponding form factors $F_{1,2}(r)$.
Here, ${\bf S}_{1,2}$ and ${\bf L}$ are the spin operators of
diquark and antidiquark and the operator of the relative orbital
angular momentum. Since we limit our considerations to the ground
states of heavy diquark-antidiquark bound systems ($\langle{\bf
L}^2\rangle=0$), the spin-orbit and tensor terms in the potential
(\ref{eq:pot}) do not contribute in the further analysis. In the
following we choose the total chromomagnetic moment of the
axial-vector diquark $\mu_d=0$. Such a choice appears to be
natural, since the long-range chromomagnetic interaction of
diquarks proportional to $\mu_d$ then also vanishes in accordance
with the flux-tube model.

\begin{table}
 \caption{Masses $M$ of heavy-diquark ($QQ'$)--light-antidiquark ($\bar q\bar q$)
 states. $T$ is the lowest threshold for decays into two
 heavy-light ($Q\bar q$) mesons and $\Delta=M-T$. All values are
 given~in~MeV.}
 \label{tab:QQmass}
\begin{ruledtabular}
\begin{tabular}{ccccrccrccr}
System&State&
\multicolumn{3}{l}{\underline{\hspace{1.cm}$Q=Q'=c$\hspace{1.cm}}}\hspace{-1.cm}&
\multicolumn{3}{l}{\underline{\hspace{1.cm}$Q=Q'=b$\hspace{1.cm}}}\hspace{-1.cm}&
\multicolumn{3}{l}{\underline{\hspace{0.6cm}$Q=c$,
$Q'=b$\hspace{0.6cm}}} \hspace{-1.5cm} \\ &$I(J^{P})$ & $M$ &$T$ &
$\Delta$ & $M$ &$T$ & $\Delta$ & $M$ &$T$ & $\Delta$\\ \hline
$(QQ')(\bar u\bar d)$\\ &$0(0^+)$ & & & & & & & 7239 & 7144 & 95\\
&$0(1^{+})$ & 3935 & 3871 & 64 & 10502 & 10604 & $-102$ & 7246 &
7190 & 56\\ &$1(1^+)$ & & & & & & & 7403 & 7190 & 213\\
&$1(0^{+})$ & 4056 & 3729 & 327& 10648 & 10558 & 90 & 7383 & 7144
& 239\\ &$1(1^{+})$ & 4079 & 3871 & 208& 10657 & 10604 & 53 & 7396
& 7190 & 206\\ &$1(2^{+})$ & 4118 & 4014 & 104& 10673 & 10650 & 23
& 7422 & 7332 & 90\\ $(QQ')(\bar u\bar s)$\\ &$\frac12(0^+)$ & & &
& & & & 7444 & 7232 & 212\\ &$\frac12(1^{+})$ & 4143 & 3975 & 168
& 10706 & 10693 & 13 & 7451 & 7277 & 174 \\ &$\frac12(1^{+})$ & &
& & & & & 7555 & 7277 & 278\\ &$\frac12(0^{+})$ & 4221 & 3833 &
388 & 10802 & 10649 & 153 & 7540 & 7232 & 308 \\ &$\frac12(1^{+})$
& 4239 & 3975 & 264 & 10809 & 10693 & 116 & 7552 & 7277 & 275 \\
&$\frac12(2^{+})$ & 4271 & 4119 & 152 & 10823 & 10742 & 81 & 7572
& 7420& 152\\ $(QQ')(\bar s\bar s)$\\ &$0(1^{+})$ & & & & & & &
7684 & 7381 & 303\\ &$0(0^{+})$ & 4359 & 3936 & 423 & 10932 &
10739 & 193 & 7673 & 7336 & 337\\ &$0(1^{+})$ & 4375 & 4080 & 295
& 10939 & 10786 & 153 & 7683 & 7381 & 302\\ &$0(2^{+})$ & 4402 &
4224 & 178 & 10950 & 10833 & 117 & 7701 & 7525 & 176\\
 \end{tabular}
\end{ruledtabular}
\end{table}

We substitute the quasipotential (\ref{eq:pot}) in the
quasipotential equation (\ref{quas}) and solve the resulting
differential equation numerically. The calculated masses $M$ of
tetraquarks with open charm and/or bottom composed from the heavy
diquark, containing two heavy quarks ($QQ'$, $Q=b,c$), and the
light antidiquark ($\bar q\bar q'$, $q=u,d,s$) are presented in
Table~\ref{tab:QQmass}. In this table we give the values of the
lowest thresholds $T$ for decays into two corresponding
heavy-light mesons [$(Q\bar
q)=D^{(*)},D_s^{(*)},B^{(*)},B_s^{(*)}$], which were calculated
using the measured masses of these mesons \cite{pdg}. We also show
values of the difference of the tetraquark and threshold masses
$\Delta=M-T$. If this quantity is negative, then the tetraquark
lies below the threshold of the decay into mesons with open
flavour and thus should be a narrow state which can be detected
experimentally. The states with small positive values of $\Delta$
could be also observed as resonances, since their decay rates will
be suppressed by the phase space. All other states are expected to
be very broad and thus unobservable. We find that the only
tetraquark which lies considerably below threshold is the $0(1^+)$
state of $(bb)(\bar u\bar d)$. All other $(QQ')(\bar q \bar q')$
tetraquarks are predicted to lie either close to or significantly
above corresponding thresholds. It is evident from the results
presented in Table~\ref{tab:QQmass} that the heavy tetraquarks
have increasing chances to be below the open flavour threshold and
thus have a narrow width with the increase of the ratio of the
heavy diquark mass to the light antidiquark mass.

\begin{table}
 \caption{Masses $M$ of diquark ($cq'$)--antidiquark ($\bar b\bar q$)
 states. $T$ is the lowest threshold for decays into two
 heavy-light ($Q\bar q$) mesons and $\Delta=M-T$; $T'$ is the
 threshold for decays into the $B_c^{(*)}$ and a light meson ($q'\bar q$), and
 $\Delta'=M-T'$. All values are
 given in MeV.}
 \label{tab:cqbqmass}
\begin{ruledtabular}
\begin{tabular}{ccccrccccrcr}
System&State&
\multicolumn{5}{l}{\underline{\hspace{2.5cm}$q'=u$\hspace{2.5cm}}}\hspace{-2.cm}&
\multicolumn{5}{l}{\underline{\hspace{2.5cm}$q'=s$\hspace{2.5cm}}}\hspace{-2.cm}
\\
&$J^{P}$ & $M$ &$T$ & $\Delta$&$T'$ & $\Delta'$ & $M$ &$T$ &
$\Delta$&$T'$ & $\Delta'$ \\ \hline $(cq')(\bar b\bar u)$\\ &$0^+$
& 7177 & 7144 & 33& 6818&359 & 7294 & 7232 & 62& 6768 & 526 \\
&$1^{+}$ & 7198 & 7190 & 8 & 6880 & 318 & 7317 & 7277 & 40 & 6820
& 497\\ &$1^+$ & 7242 & 7190 & 52 & 6880 & 362 & 7362 & 7277 & 85
& 6820 & 542\\ &$0^{+}$ & 7221 & 7144 & 77 & 6818& 403 & 7343 &
7232 & 111 & 6768 & 575\\ &$1^{+}$ & 7242 & 7190 & 52 & 6880 & 362
& 7364 & 7277 & 87 & 6820 & 544\\ &$2^{+}$ & 7288 & 7332 & $-44$
&7125 &163 & 7406 & 7420 & $-14$ & 7228 & 178\\ $(cq')(\bar b\bar
s)$\\ &$0^+$ & 7282 & 7247 & 35 & 6768 &514 & 7398 & 7336 & 62&
6818& 580 \\ &$1^{+}$ & 7302 & 7293 & 9 & 6820& 482 & 7418 & 7381
& 37 &6880 & 538\\ &$1^+$ & 7346 & 7293 & 53 & 6820& 526 & 7465 &
7381 & 84 & 6880 & 585\\ &$0^{+}$ & 7325 & 7247 & 78 & 6768 & 557
& 7445 & 7336 & 109& 6818 & 627\\ &$1^{+}$ & 7345 & 7293 & 52&
6820& 525& 7465 & 7381 & 84& 6880& 585 \\ &$2^{+}$ & 7389 & 7437 &
$-48$& 7228& 161& 7506 & 7525 & $-19$& 7352 & 154\\
 \end{tabular}
\end{ruledtabular}
\end{table}

In Table~\ref{tab:cqbqmass} the calculated masses $M$ of
tetraquarks composed from a $(cq)$ diquark and a $(\bar b\bar q)$
antidiquark are listed.~\footnote{Such $(cu)(\bar b\bar u)$
tetraquarks were recently argued \cite{lipkin} to be the best
candidates for experimental detection.} We also give the lowest
thresholds $T$ for decays into heavy-light mesons as well as
thresholds $T'$ for decays into the $B_c^{(*)}$ and light ($q'\bar
q$) mesons and $\Delta^{(')}=M-T^{(')}$.\footnote{For the
non-strange $(cq)(\bar b\bar q)$ tetraquarks we give thresholds
$T'$ for decays of the $I=0$ states into $B_c^{(*)}$ and $\eta$ or
$\omega$. These states should be more stable than the $I=1$ ones,
since their decays to $B_c^{(*)}$ and
 $\pi$ violate isospin.} We find that only $2^+$ states of $(cq')(\bar
b \bar q)$ have negative values of $\Delta$ and thus they should
be stable with respect to decays into heavy-light ($B$ and $D$)
mesons. The predicted masses of lowest $1^+$ states of $(cu)(\bar
b \bar u)$ and $(cu)(\bar b \bar s)$ tetraquarks lie only slightly
above the corresponding thresholds $T$. However, all $(cq)(\bar
b\bar q)$ tetraquarks are found to be significantly above the
thresholds $T'$ for decays into the $B_c^{(*)}$ and light ($q'\bar
q$) mesons. Nevertheless, the wave function of the spatially
extended $(cq)(\bar b\bar q)$ tetraquark would have little overlap
with the wave function of the compact $B_c$ meson
\cite{torn,swanson}, thus substantially suppressing the decay rate
in this channel. Therefore the above-mentioned $(cq)(\bar b\bar
q)$ tetraquark states which are below the $BD$ threshold have good
chances to be rather narrow and could be detected experimentally.

It is important to note that the comparison of the masses of heavy
tetraquarks given in Tables~\ref{tab:QQmass} and
\ref{tab:cqbqmass} with our previous predictions \cite{efght}
shows that the $(QQ')(\bar q\bar q')$ states are, in general,
heavier than the corresponding $(Qq)(\bar Q'\bar q')$ ones. This
result has the following explanation. Although the relation
\cite{nl} $M_{QQ}+M^S_{qq}\le 2M_{Qq}$ holds between diquark
masses, the binding energy in the heavy-light diquark ($Qq$)--
heavy-light antidiquark ($\bar Q\bar q$) bound system is
significantly larger than in the corresponding heavy diquark
($QQ$)--light antidiquark ($\bar q\bar q$) one. This fact is well
known from the meson spectroscopy, where heavy quarkonia $Q\bar Q$
are more tightly bound than heavy-light mesons $Q\bar q$. For
instance, we found that some of the $(cu)(\bar c\bar u)$
tetraquarks lie below open charm thresholds while all ground-state
$(cc)(\bar u \bar d)$ tetraquarks are found to be above such
thresholds.

\begin{table}
 \caption{Comparison of different theoretical predictions for the masses
 of heavy $(QQ')(\bar q \bar q')$ tetraquarks (in MeV).}
 \label{tab:compmass}
\begin{ruledtabular}
\begin{tabular}{cccccccccc}
System&$I(J^{P})$& this work& \cite{sbs} &\cite{jr} &\cite{bs}
&\cite{vfvs} &\cite{vvt}&\cite{gn} &\cite{nn}\\ \hline $(cc)(\bar
u\bar d)$\\ &$0(1^{+})$ & 3935 & 3931 & 3876 & & 3764 & 3927&3905
& $4000\pm200$ \\ &$1(0^{+})$ & 4056 & & & & 4150 & &\\
&$1(1^{+})$ & 4079 & & & & 4186 & &\\ &$1(2^{+})$ & 4118 & & & &
4211 & &\\ $(bb)(\bar u\bar d)$\\ &$0(1^{+})$ & 10502 & 10525 &
10504 & 10558 & 10261 & 10426& & $10200\pm300$ \\ &$1(0^{+})$ &
10648 & & 10587 & 10766 & 10690 & &\\ &$1(1^{+})$ & 10657 & 10712
& 10644 & 10774 & 10698 & &\\ &$1(2^{+})$ & 10673 & 10735 & &
10790 & 10707 & &\\ $(bb)(\bar u\bar s)$\\ &$\frac12(1^{+})$ &
10706 & 10680 & & & & &\\ &$\frac12(2^{+})$ & 10823 & 10816 & & &
& &\\ $(bc)(\bar u\bar d)$\\ &$0(0^{+})$ & 7239 & 7206 & & & & &\\
&$0(1^{+})$ & 7246 & 7244 & & & & &\\ &$1(2^{+})$ & 7422 & 7422 &
& & & &\\ $(bc)(\bar u\bar s)$\\ &$\frac12(2^{+})$ & 7572 & 7496 &
& & & &\\
\end{tabular}
\end{ruledtabular}
\end{table}

In Table~\ref{tab:compmass} we confront our results for masses of
heavy $(QQ')(\bar q \bar q')$ tetraquarks with other theoretical
predictions \cite{sbs,jr,bs,vfvs,vvt,gn,nn}. In Ref.~\cite{sbs}
the authors solve the four-body problem using the expansion in the
harmonic-oscillator basis in the framework of the nonrelativistic
quark model with a phenomenological potential. The same model with
a different expansion basis, which can accommodate asymptotic
states of two free mesons, is applied for the calculation of
heavy-tetraquark properties in Ref.~\cite{jr}. The stability of
tetraquarks with heavy flavours is studied by using a variational
approach and a nonrelativistic potential model in Ref.~\cite{bs}.
In Refs.~\cite{vfvs,vvt} tetraquarks are analyzed in the chiral
constituent quark model using a variational approach. The
potential of this model includes one-gluon, confinement and
meson-exchange interactions. The existence of a virtual tetraquark
state $cc\bar u\bar d$ is discussed in Ref.~\cite{gn} on the basis
of semi-empirical mass relations. QCD sum rules are applied for
the $QQ\bar u\bar d$ tetraquarks in Ref.~\cite{nn}. The main
difference between our approach and the above quoted papers
consists in 
that, from the very beginning, we explicitly reduce the
relativistic four-body problem to the subsequent solution of two
relativistic two-body problems assuming the diquark-antidiquark
structure of the $(QQ')(\bar q\bar q')$ tetraquarks. From
Table~\ref{tab:compmass} we see that most of the presented
approaches predict that only the $1^+$ state of the $(bb)(\bar
u\bar d)$ tetraquark lies below the open-bottom threshold (see
also \cite{lpk}). In Ref.~\cite{jr} it is claimed that also the
$1^+$ state of the $(cc)(\bar u\bar d)$ tetraquark is weakly bound
against the $DD^*$ threshold, if it has a molecular structure.
Note that such structures are absent in our approach. A large
binding energy in the $1^+$ state of the $(cc)(\bar u\bar d)$
tetraquark is found only in Ref.~\cite{vfvs} and it is claimed to
originate from the meson-exchange part of the quark interaction
potential. The recent QCD sum rule analysis \cite{nn} finds that
only the $(bb)(\bar u\bar d)$ tetraquark is expected to be a
narrow state.

\section{Conclusions}
\label{sec:concl}

In this paper we have calculated the masses of the ground states
of tetraquarks with two heavy quarks assuming the
diquark--antidiquark structure. Such approximation allowed us to
reduce the very complicated relativistic four-body problem to the
solution of two --- significantly more simple --- relativistic
two-body problems. All considerations were done in the framework
of the relativistic quark model which proved to be successful in
describing numerous properties of mesons and baryons
\cite{hbar,efght,efgm,exhbar,efg,egf,fg,lmes}. The parameters of
the model were fixed previously from the meson sector and are kept
unchanged in the present analysis. The diquarks were treated
dynamically. Their masses and form factors, which take into
account the diquark structure, were calculated on the basis of a
numerical solution of the corresponding relativistic
quasipotential equation. Note that they are the same as in our
previous studies of light and heavy diquarks in heavy \cite{hbar}
and doubly-heavy baryons \cite{efgm}, respectively. Light quarks
and diquarks were treated completely relativistically without
applying the $v/c$ expansion. It was found that both the
relativistic dynamics of light diquarks as well as their internal
structure play an important role in the description of
diquark-antidiquark bound states. The binding of a heavy diquark
and a light antidiquark turned out to be weaker than the binding
of a corresponding heavy-light diquark and heavy-light
antidiquark. Thus, in contrast to the $(cq)(\bar c\bar q)$
tetraquarks, which were discussed previously \cite{efght}, all the
$(cc)(\bar q\bar q')$ tetraquarks are predicted to be above the
decay threshold into the open charm mesons. Only the
$I(J^P)=0(1^+)$ state of $(bb)(\bar u\bar d)$ was found to lie
below the $BB^*$ threshold. As a result, this state can decay only
weakly and thus it should be narrow. The strange partner of this
state $(bb)(\bar u\bar s)$ is predicted to lie slightly (13 MeV)
above the $B^*B_s$ threshold and, in principle, could be observed
as a not too broad resonance decaying mainly into $B^*B_s$.

The investigation of the decay widths of heavy tetraquarks which are
predicted to lie below the threshold of the open flavours represents another very
important and interesting problem. It could be considered by means of
the relativistic generalization of the analysis performed in
Ref.~\cite{Melikhov:2006ec}. However, this problem is beyond the scope
of the present paper and will be considered elsewhere.

The authors are grateful to V.~A.~Matveev, D.~Melikhov,
M.~M\"uller-Preussker and V.~I.\ Savrin for support and useful
discussions. Two of us (R.N.F. and V.O.G.) were supported in part
by the {\it Deutsche Forschungsgemeinschaft} under contract Eb
139/2-4 and by the {\it Russian Foundation for Basic Research}
under Grant No.05-02-16243.

\end{document}